\documentclass[prl,preprint,aps,superscriptaddress]{revtex4-2}

\usepackage[T1]{fontenc}
\usepackage{enumerate}
\usepackage{times}
\usepackage{amssymb}
\usepackage{xcolor}
\usepackage{graphicx}

\newcommand{\ket}[1]{\left|#1\right\rangle}
\newcommand{\bra}[1]{\left\langle #1\right|}

\newcommand{\expectation}[3]{\left\langle #1\left|#2\right|#3\right\rangle}

\begin{document}

\title{Overcoming Stark-Shift Constraints in Phase-Controlled Rydberg Two-Qubit Gates}  

\author{Ignacio R. Sola}
 \affiliation{Departamento de Quimica Fisica, Universidad Complutense de Madrid, 28040 Madrid, Spain} 
\author{Sebastian C. Carrasco}
\affiliation{DEVCOM Army Research Laboratory, 2800 Powder Mill Road, Adelphi, Maryland 20783, USA}
\author{Vladimir~S.~Malinovsky}
\affiliation{DEVCOM Army Research Laboratory, 2800 Powder Mill Road, Adelphi, Maryland 20783, USA}
\author{Seokmin Shin}
\affiliation{School of Chemistry, Seoul National University, 08826 Seoul, Republic of Korea}
\author{Bo Y. Chang}
\email{Contact author: boyoung@snu.ac.kr}
 \affiliation{School of Chemistry, Seoul National University, 08826 Seoul, Republic of Korea}
  
\begin{abstract}
Stark shifts introduce additional phases that constrain the set of entangling gates that can be prepared via two-photon transitions in the strong Rydberg blockade limit. For non-independently addressed qubits, by controlling the absolute phases and the local amplitudes of the pulses at each qubit, we show that any two-qubit phase gate can be prepared with high fidelity using a three-pulse sequence. Based on these insights, we introduce two robust control schemes tailored to different phase gates that yield better results with pulse sequences of either even or odd length.

\end{abstract}

\maketitle

\section{Introduction}

The quest for scalable quantum computation has progressed from fundamental research in academic laboratories to the development of commercial prototypes, with several physical platforms -superconducting circuits~\cite{Kjaergaard_ARConMatPhy2020}, trapped ions~\cite{Wineland_Nat2008}, photonic systems~\cite{OBrien_Science2007}, semiconductor devices~\cite{Zwanenburg_RMP2013}, and neutral atoms~\cite{Saffman_RMP2010}- now under active competition. Despite these advances, there remains considerable room at the research level to explore unconventional approaches that may ultimately offer advantages in scalability, fidelity, or architectural simplicity. 
Neutral atoms trapped in optical lattices or optical tweezers~\cite{Browaeys_PRX2014, Browaeys_Science2016, Wilson_PRL2022, Thompson_PRXQ2022, Ahn_OptE2016} 
are a versatile resource for implementing 
multi-particle entanglement~\cite{Lukin_PRL2018, Zhan_PRL2017, Ahn_PRL2020, Grangier_PRL2010, Saffman_Nature2022, Saffman_PRA2015, Saffman_PRA2010, Picken_QCT2018, Malinovsky_PRA2004, Malinovsky_PRL2004, Malinovsky_PRL2006}, simple quantum circuits~\cite{Jaksch_PRL2000, Saffman_QIP2011, Saffman_PRA2015, Lukin_PRL2001, Lukin_PRL2019, Cohen_PRXQ2021, Shi_QSTech2022, Shi_PRApp2018, Adams_PRL2014, Adams_JPB2019, Malinovsky_PRA2014, Goerz_JPB2011, Morgado_AVSQSci2021, Alexey_PRL2021, Sanders_PRA2020} and even
quantum gates across different quantum computing platforms~\cite{Hennrich_Nature2020, Molmer_PRX2020, Khazali_OE2023, Pohl_Quantum2022, Raithel_PRL2011, Crane_PRR2021}.
Since the early proposals of Jaksch et al.~\cite{Jaksch_PRL2000, Deutsch_FP2000, Mompart_PRL2003, Saffman_PRA2005}, neutral atoms have been seen as a promising platform to prepare entangling gates via the strong dipole blockade mechanism.

Historically, relatively slow two-qubit gate times and modest fidelities limited the competitiveness of Rydberg-mediated entangling operations~\cite{Grangier_PRL2010,Isenhower_PRL2010}, but recent technological breakthroughs have changed this landscape: stronger and more controllable Rydberg–Rydberg interactions~\cite{Browaeys_NP2020}, improved trapping and rearrangement techniques enabling near-deterministic loading~\cite{Endres_Science2016,Browaeys_Science2016,Anh_NatCom16}, and substantially longer coherence times~\cite{Madjarov_NatPhys2020} have together renewed interest in neutral-atom quantum computation. These developments have motivated a new wave of research into simpler, more robust, and more reliable entangling-gate protocols~\cite{Graham_PRL2019,Lukin_PRL2019,Carrasco_arXiv2025,Cole_arxiv2025}, and the use of optimization techniques~\cite{Goerz_JPB2011,Goerz_NJP2014,Goerz_PRA2014,Jandura_Quant2022}. Our work contributes to this effort by exploring non-standard control strategies that rely on precise manipulation of the spatial and temporal characteristics of the optical pulses driving the qubits~\cite{Sola_Nanoscale2023,Sola_PRA2023,Sola_AIPadv2023,Sola_PRA2024}.

Building on the paradigmatic Jaksch protocol (JP) for implementing a controlled-Z (CZ) gate based on a sequence of $\pi-2\pi-\pi$ pulses~\cite{Jaksch_PRL2000}, we recently proposed a minimal modification designed to operate effectively with closely spaced 
qubits, that are not fully distinguishable -the pulses cannot act independently on each qubit- nor fully indistinguishable -the pulses do not necessarily act on both qubits in the same way. 
We termed our scheme the Symmetrically Orthogonal Protocol (SOP).
There is, then, an opportunity to arrange the qubits and pulses such as to control the local amplitude of each pulse on each qubit. 
The spatial configuration of the driving fields is chosen such that, after the first pulse, the system—when initialized in the $\ket{11}$
state -occupies a superposition of $\ket{1r}$ and $\ket{r1}$ states that is dark to the second pulse. Consequently, the second pulse leaves the wave function unaffected, and the third pulse reverses the excitation, restoring the system to $\ket{11}$ with the required phase shift. They are called {\em spatially orthogonal} conditions.
The analysis of the gate, as in other publications~\cite{Jaksch_PRL2000,Sola_Nanoscale2023} considered single-photon excitation of the Rydberg state.

In most experimental setups, however, the excitation of the high-lying Rydberg state is done via nonresonant two-photon absorption~\cite{Saffman_RMP2010,Grangier_Nphys2009}.
In the present work, 
using the adiabatic elimination approximation, we show that
the effective two-photon Hamiltonian introduces additional dynamical phases arising from Stark shifts associated with the intermediate off-resonant level~\cite{Molmer_PRA2008,Petrosyan_PRA2017},
that prevent the preparation of certain two-qubit phase gates.
For instance, we explicitly demonstrate that for two-photon transitions, the JP and SOP cannot be used to prepare a two-qubit phase where all the signs of the amplitudes in the computational states other than $\ket{00}$ at final time are flipped~\cite{Jaksch_PRL2000,Goerz_JPB2011,Sola_Nanoscale2023}.
However, by controlling the absolute optical phases of the driving pulses, we show that any entangling gate can be realized. 
Moreover, such phase control significantly broadens the range of values of pulse areas that yield high-fidelity performance.

Finally, motivated by these insights, we propose extensions of the SOP protocol that operate with pulse sequences with different number of pulses, with restrictions such that all odd-numbered pulses in the sequence have the same areas and spatial parameters, and all even-numbered pulses in the sequence have also the same areas and spatial parameters.
By lifting the spatially orthogonal conditions,
we propose more efficient extensions of the SOP tailored for systems with non-independently addressed qubits, specifically for pulse sequences containing an even rather than an odd number of pulses -thereby complementing the original SOP and enabling new families of robust entangling operations.

\section{Qubit set-up}

We consider the usual setup of neutral atoms trapped by optical tweezers,
where the qubit states $\ket{0}$ and $\ket{1}$ are encoded in hyperfine levels of the ground state.
Two-qubit phase gates, such as the CZ gate, require the population to remain in the starting qubit state, gaining some phase in the process such that the final state can be entangled.
These gates can be prepared by driving the population through
a Rydberg state, thus making use of the dipole blockade mechanism, which
bans the excitation of more than one atom in a Rydberg state when the atoms are
close enough, within the so-called dipole blockade radius, $R_{\cal B}$.
The laser frequencies are chosen such that one of the qubits, typically $\ket{00}$, is very off-resonant, and is thus not affected by the protocol.
In single-photon transitions between two states, $\ket{1}$ and the Rydberg state $\ket{r}$, driven by a resonant pulse with envelope $E(t)$, the time-dependent Schr\"odinger equation (TDSE) can be solved analytically, with the result
\begin{equation}
    \ket{\Psi(t)} = \cos(\theta(t)) \ket{1} + i\sin(\theta(t)) \ket{r}
\end{equation}
where the mixing angle 
\begin{equation}
    \theta(t) = \frac{1}{2} \int_0^t \Omega(t') dt'
\end{equation}
with Rabi frequency $\Omega(t) = \mu E(t)$, where $\mu$ is the transition dipole moment.
The resonant transition will induce full population transfer from $\ket{1}$ to $\ket{r}$, or full population return to $\ket{1}$, depending on the pulse area $S$, defined as the 
integral of the Rabi frequency at the end of the pulse, $\Omega(t)$ ($S = 2\theta(\infty)$). 
Hence, a pulse of area $\pi$, known as a $\pi$-pulse, will prepare the state $\ket{r}$, or more precisely
$U_\pi\ket{1} = i\ket{r} \equiv e^{i\pi/2}\ket{r}$, where the transition gains a $\pi/2$ phase and we used the time-evolution operator $U$ to formally represent the solution of the time-dependent Schr\"odinger equation, indicating the pulse area with a subscript.
A $2\pi$-pulse, in contrast, will prepare 
$U_{2\pi}\ket{1} = -\ket{1} = e^{i\pi}\ket{1}$.
On the other hand, if $\ket{0}$ is far off-resonance, we can guarantee that, for any pulse area, $U\ket{0} = \ket{0}$.

Next, we consider two trapped atoms at a distance large enough that we can drive each atom by a different field, without affecting the other.
The possible qubit states of the system (the computational basis states) are $\ket{00}, \ket{01}, \ket{10},\ket{11}$, where we assume that the first (to the left) is qubit $1$ and the second is qubit $2$. To simplify the notation, we add a parenthesis to the time-evolution operator indicating the qubit to which the pulse is acting.
In the original protocol proposed by Jaksch et al.,
a $\pi$ pulse acting on the first qubit, is followed by a $2\pi$ pulse acting on the
second qubit, followed by a repetition of the first step.
As long as the pulses do not overlap in time, one can analytically predict the result of the dynamics acting sequentially with the time-evolution operators, resulting in
\begin{eqnarray}
U_{\pi}(1)U_{2\pi}(2)U_{\pi}(1) = \mathrm{diag}(1,-1,-1,-1) \equiv {\cal C}^-
\end{eqnarray}
where $\mathrm{diag}(1,-1,-1,-1)$ is a diagonal matrix with diagonal elements as shown in parenthesis.
In the following, we refer to this version of the CZ gate as the ${\cal C}^-$ gate.
The key step in the protocol is the suppression of the second Rydberg state 
through dipole blockade, explicit in the process 
$U_{\pi}(1)U_{2\pi}(2)U_{\pi}(1) |11\rangle = i U_{\pi}(1)U_{2\pi}(2) |r1\rangle = i U_{\pi}(1) |r1\rangle$, where the second pulse apparently does not act, which then leads to $-\ket{11}$.
Within the paradigm of the JP, to operate with fast gates the Rydberg blockade must be strong, which can be met by two
conditions: exciting high Rydberg states and arranging the atoms at small internuclear distances.
The first condition is satisfied by using two laser pulses that drive the population non-resonantly through an intermediate step.
We cannot assume that the transition from $\ket{1}$ to $\ket{r}$ can be single-photon.
Hence, each pulse in the three-step protocol must be replaced
by a pair of so-called pump and Stokes pulses. 
The second condition is limited by
our ability to address the qubits independently, as in Jaksch protocol.
This condition is overriden in the SOP.
Fig.~1 shows a diagram of the energy levels in the qubit setup, including the computational basis, the Rydberg state, and the intermediate state $\ket{i}$ (typically an nP atomic state).

Although it is possible to drive resonantly two-photon transitions avoiding 
population in the intermediate state~\cite{Carrasco_PRL2025}, 
we will assume in this work that the two-photon transition is nonresonant, 
as in most laboratory implementations.
To obtain analytical formulas, we will work in the far offresonant limit. 
This allows us 
to work with an effective two-level Hamiltonian after adiabatic elimination of the amplitude in $\ket{i}$
to study the two-photon resonant transition from state $|1\rangle$ 
to state $|r\rangle$.
In the rotating-wave approximation,\\
\begin{equation}
H(t) = -\frac{1}{2}
\left( \begin{array}{cc} \frac{\displaystyle \Omega^2_{P}(t)}
{\displaystyle 2\Delta} &
 \frac{\displaystyle \Omega_{P}(t) \Omega_{S}(t)}{\displaystyle 2\Delta} e^{i\phi} \\ \\
 \frac{\displaystyle \Omega_{P}(t) \Omega_{S}(t)}{\displaystyle 2\Delta} e^{-i\phi}
 & \frac{\displaystyle \Omega^2_{S}(t) }{\displaystyle 2\Delta} \end{array} \right)
\label{H1q2l}
\end{equation}
where $\phi = \phi_{S} - \phi_{P}$ 
is the phase difference between the phase-locked pump and Stokes pulses and
$\Omega_P = \mu_{0r} E_P(t)/\hbar$, $\Omega_S = \mu_{0r} E_S(t)/\hbar$,
where $E_P(t)$ and $E_S(t)$ are the pump and Stokes pulse envelopes.
For coincident pulses, 
$\Omega_{P}(t) = \Omega_{S}(t) = \Omega_{0}(t)$, the TDSE equation can be solved analytically, with time evolution operator given by
\begin{equation}
U(t) = e^{ i\theta(t)} \left( \begin{array}{cc} \cos \theta(t) & i e^{i\phi} \sin \theta(t) 
\\ i e^{-i\phi} \sin \theta(t) & \cos \theta(t) \end{array} \right)
\label{U1d2l}
\end{equation}
where the mixing angle is now
\begin{equation}
\theta(t) = 
\int_{-\infty}^{t} \frac{\Omega^2_{0}(t')}{4\Delta} dt' =
 \frac{1}{2} \int_{-\infty}^{t}\tilde{\Omega}(t') dt'
\end{equation}
and where we have defined the Rabi frequency as
$\tilde{\Omega}(t) = \Omega^2_{0}(t)/2\Delta$,
with a tilde to remind us that this is an effective two-photon Rabi frequency.
Its integral during the pulse duration will be the pulse area $S$.
The Stark-shifts in the Hamiltonian [Eq.(\ref{H1q2l})] generate a global phase shift
in the time-evolution operator [Eq.(\ref{U1d2l})], such that while for a typical (single-photon) two-level Hamiltonian, a $2\pi$-pulse changes the state from 
$|1\rangle$ to $\cos\pi|1\rangle = -|1\rangle$ (and a $\pi$ pulse moves the population
from $|1\rangle$ to $i|r\rangle$), 
the same pulse with the Stark shifts,  
operates as $e^{iA/2} \cos(A/2)\ket{1} = e^{i\pi} \cos(\pi)\ket{1} = \ket{1}$, 
thus inducing no change in the phase of the state at the end of the process.
Therefore, using the JP with two-photon transitions in adiabatic elimination conditions prepares the gate
\begin{eqnarray}
U_{\pi}(1)U_{2\pi}(2)U_{\pi}(1) = \mathrm{diag}(1,1,1,-1) \equiv {\cal C}^+
\end{eqnarray}
This gate, like ${\cal C}^-$, can be used to generate entanglement. By adding local operations (single-qubit gates) that act on each qubit independently one can switch from one gate to the other.
However, it is interesting to understand why some two-qubit phase gates are simpler to obtain than others and to propose 
protocols that implement each gate via a simple pulse sequence, such as the JP.
Alternatively, one can seek phase-gate protocols that generate entanglement, regardless of the specific signature.
They can be evaluated by computing the entangling power.


\section{The two-photon Symmetric Orthogonal Protocol}

\begin{figure}
\includegraphics[width=8cm]{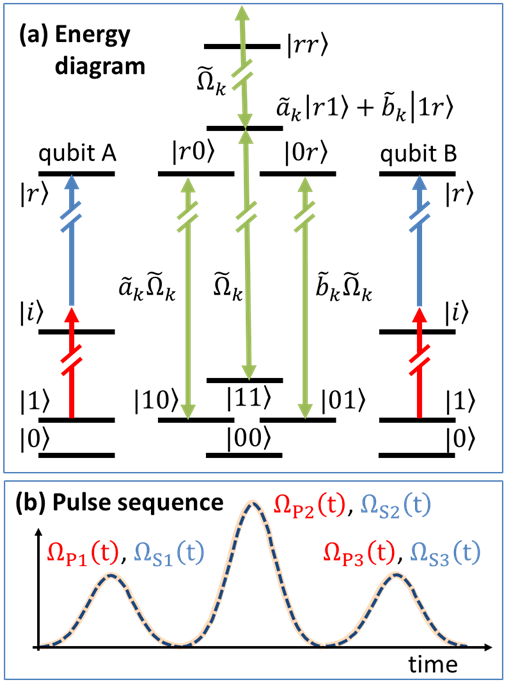}
\caption{(a) Diagram of the energy levels as the qubits come close and we describe the two-photon interaction in the adiabatic approximation, with pulses overlapping both qubits with different factors $a_k$ and $b_k$. (b) Symmetrical $3$-(double)-pulse sequence where the pump and Stokes Rabi frequencies are always identical and the pulse area of the first and third pulses in the sequence are the same and half the pulse areas of the second pulses.} 
\label{fig1}
\end{figure}

As previously discussed, to enhance the dipole blockade 
and speed up the two-qubit gates in the JP 
can be achieved by decreasing the distance between the qubits from
typical values (on most platforms) around $5\mu$m to distances as short as possible ($\sim 1\mu$m). Then the qubits cannot be addressed independently.
Therefore, we will now extend the Hamiltonian and equations of motion to describe such situation.
By design, 
the energy difference between the computational basis states is large enough that the Hamiltonian is block-diagonal.
As a result, the evolution can be analyzed independently for each 
state $|00\rangle$, $|01\rangle$, $|10\rangle$, $|11\rangle$.
When the dynamics is initiated in $|11\rangle$, 
the relevant subsystem forms a three-level system in a $V$ configuration (although one can defined a suitable basis to further reduce the system to a two-level system, specific for each qubits and fields arrangement, as indicated in Fig.~1[a]).
For $|01\rangle$ and $|10\rangle$, the evolution reduces to two-level dynamics, while the state $|00\rangle$ remains decoupled.
In Fig.~1 we show a diagram of the relevant states before and after adiabatic elimination of the intermediate states $\ket{i}$.

In the SOP scheme, we use structured light or a superposition of Gaussian beams, so that we control the local amplitude of the fields at each qubit location~\cite{Sola_Nanoscale2023,Sola_PRA2024}.
Hence, we define spatial or {\em geometrical factors} $a_P$ and $a_S$,
which multiply the {\em local} Rabi frequency of each field ($\Omega_P$ or $\Omega_S$) acting on qubit $a$. Similarly,
$b_P$ and $b_S$ are the corresponding coefficients for qubit $b$.
As before, to simplify the notation, we call 
\begin{equation}
\tilde{\Omega}_{\alpha\beta}(t) = \Omega_i(t) \Omega_j(t) / 2\Delta
  \hspace{0.5cm} \text{(with $\alpha, \beta = P,S$)}  
\end{equation}
the effective two-photon Rabi frequencies.
This will be valid for each step in the gate sequence. 
Later on we will need to attach an additional label referring to the specific step in the pulse sequence to these effective two-photon Rabi frequencies. 

After adiabatic elimination of the intermediate state and assuming strong dipole blockade, the effective Hamiltonian for the system starting in the $|11\rangle$ state can be written, in the ordered basis
$\lbrace |11\rangle, |r1\rangle, |1r\rangle \rbrace$, as
\begin{equation}
H^V(t) = -\frac{1}{2}
\left( \begin{array}{ccc} \left( 
a^2_P + b_P^2 \right) \tilde{\Omega}_{PP}(t) & a_P a_S  \tilde{\Omega}_{PS}(t) e^{i\phi} & b_P b_S \tilde{\Omega}_{PS}(t) e^{i\phi} \\
a_P a_S  \tilde{\Omega}_{PS}(t) e^{-i\phi} & b_P^2 \tilde{\Omega}_{PP}(t) + a_S^2 \tilde{\Omega}_{SS}(t)  & 0 \\
b_P b_S  \tilde{\Omega}_{PS}(t) e^{-i\phi} & 0 &  a_P^2 \tilde{\Omega}_{PP}(t) + b_S^2 \tilde{\Omega}_{SS}(t)  \\
 \end{array} \right)
\label{HV}
\end{equation}
Notice that because every field acts on both qubits at the same time, the AC Stark shifts have two components. In particular, to compute the Stark shift over state $\ket{r1}$, in addition to the effect of the virtual photon that nonresonantly couples this state with the intermediate state $\ket{i1}$, given by $a_S^2 \tilde{\Omega}_{SS}(t)$, we must also include the coupling with the intermediate state at the other qubit $\ket{ri}$, given by $b_P^2 \tilde{\Omega}_{PP}(t)$. Similarly, there are two contributions to the AC Stark shift of states $\ket{1r}$ and $\ket{11}$.

The effective Hamiltonian for the two-level systems will be as equation (\ref{H1q2l}),
but we need to add the coefficients $a_P a_S$ to the
Rabi frequencies for qubit $A$, and the coefficients $b_P b_S$
for qubit $B$, as
\begin{equation}
H^A(t) = -\frac{1}{2}
\left( \begin{array}{ccc} a^2_P \tilde{\Omega}_{PP}(t) & a_P a_S  \tilde{\Omega}_{PS}(t) e^{i\phi} \\  
a_P a_S  \tilde{\Omega}_{PS}(t) e^{-i\phi} & a_S^2 \tilde{\Omega}_{SS}(t) \\ \end{array} \right)
\end{equation}
acting on states ordered as  $\lbrace |10\rangle , |r0\rangle \rbrace$, and
\begin{equation}
H^B(t) = -\frac{1}{2}
\left( \begin{array}{ccc} b^2_P \tilde{\Omega}_{PP}(t) & b_P b_S  \tilde{\Omega}_{PS}(t) e^{i\phi} \\  
b_P b_S  \tilde{\Omega}_{PS}(t) e^{-i\phi} & b_S^2 \tilde{\Omega}_{SS}(t) \\ \end{array} \right)
\end{equation}
acting on states ordered as  $\lbrace |01\rangle , |0r\rangle \rbrace$.
Finally, the full $8$-level system in our Hilbert space is completed with the
Hamiltonian $H^D = 0 \ket{00}\bra{00}$. 
Therefore, the full Hamiltonian consists of 
$H^D \oplus H^A \oplus H^B \oplus H^V$.

The SOP follows the JP strategy as closely as possible, focusing on gate protocols that use a three-(double) pulse sequence with
fully overlapping pump and Stokes pulses in each step of the sequence, as shown in Fig.~1(b). Then
$\Omega_P(t)= \Omega_S(t)$, such that $\tilde{\Omega}_{PP}(t) = \tilde{\Omega}_{SS}(t) =
\tilde{\Omega}_{PS}(t)$. 
We now add a label to identify the order of the pulse in the sequence, so in the previous
equations we substitute any $\tilde{\Omega}_{\alpha\beta}(t)$ ($\alpha, \beta = P, S$) by $\tilde{\Omega}_{\alpha\beta,k}(t)$, 
with $k = 1, 2, \ldots, M$, and since all of them are equal, we remove the $\alpha\beta$ subindex leaving it as $\tilde{\Omega}_k(t)$ to simplify the notation. 
Finally, we choose
the pump and Stokes geometrical factors such that
$a_P = a_S$ and $b_P = b_S$, which makes all diagonal terms in the Hamiltonian (\ref{HV}) equal.
This can be easily achieved if the beams or structural features of both pulses are
the same in each qubit, while they can differ with respect to the different qubits
({\em e.g.} the first laser can be more focused in qubit $A$ than in qubit $B$, so that
$a_P = a_S$ is larger than $b_P = b_S$). 
Adding a label again to identify the step in the sequence,  
$a_P a_S \rightarrow a_{P,k} a_{S,k} = a_{P,k}^2 \equiv \tilde{a}_k$ and $b_P b_S \rightarrow b_{P,k} b_{S,k} = b_{P,k}^2 \equiv \tilde{b}_k$,
where we drop the unnecessary subindex referring to the qubit, since both geometrical factors have the same value, and added a tilde to remind us that this factor is a product of two factors.
Although the Hamiltonian can be used for any
arrangement, we will typically assume that for the first laser (and the third, which is equal to the first, as shown in Fig.1[b]), $|\tilde{a}_1| \gg |\tilde{b}_1|$, while the opposite is true for the second laser.
The more general results obtained by lifting these constraints will be analyzed elsewhere.
Furthermore, without loss of generality, we will also assume that $\tilde{a}_k^2 + \tilde{b}_k^2 = 1$
for each pulse, scaling, as necessary, the Rabi frequencies.
Then the diagonal terms of the $H^V$ Hamiltonian are just $\tilde{\Omega}_k(t)$ 
and the Hamiltonian for each step in the sequence can be written as
\begin{eqnarray}
{\sf H}^V_k(t) = -\frac{1}{2} \tilde{\Omega}_k(t)
\left( \begin{array}{ccc} 1 & \tilde{a}_k   e^{i\phi_k} & \tilde{b}_k  e^{i\phi_k} \\
\tilde{a}_k   e^{-i\phi_k} & 1   & 0 \\
\tilde{b}_k   e^{-i\phi_k} & 0 &  1   \\ \end{array} \right) ,  \\
{\sf H}^T_k(t) = -\frac{1}{2} \tilde{\Omega}_k(t)
\left( \begin{array}{cc} |\alpha_k|  & \alpha_k  e^{i\phi_k} \\  
\alpha _k e^{-i\phi_k} & |\alpha_k|  \\ \end{array} \right) \hspace{0.5cm} (\alpha = \tilde{a}, \tilde{b}; T = A, B),
\end{eqnarray}
It is possible to obtain simple close-analytical forms for the time-evolution operators
under these Hamiltonians. 
At the end of the pulse, the propagator will be
\begin{equation}
U^{V}_{k} = e^{iS_k/2} 
\left( \begin{array}{ccc}
\cos (S_k/2) & i e^{i\phi_k} \tilde{a}_{k} \sin (S_k/2) & i e^{i\phi_k} \tilde{b}_k \sin (S_k/2)  \\
i e^{-i\phi_k}\tilde{a}_{k} \sin (S_k/2) & \tilde{a}_{k}^2 \cos (S_k/2) + \tilde{b}_k^2 & 
\tilde{a}_{k}\tilde{b}_{k} \left[ \cos (S_k/2) - 1 \right]  \\
i e^{-i\phi_k}\tilde{b}_{k} \sin (S_k/2) & \tilde{a}_{k}\tilde{b}_{k} \left[ \cos (S_k/2) - 1 \right]  & 
\tilde{b}_{k}^2 \cos (S_k/2) + \tilde{a}_k^2 \end{array} \right)  \label{UV}
\end{equation}
where $S_k$ is the area of the $k$-pulse.
 
For the two-level Hamiltonians, we obtain
\begin{equation}
U^T_{k} = e^{i\alpha_k S_k/2} 
\left( \begin{array}{cc}
\cos (\alpha_k S_k/2) & i e^{i\phi_k} \sin (\alpha_k S_k/2) \\
i e^{-i\phi_k} \sin (\alpha_k S_k/2) & \cos (\alpha_k S_k/2)  \end{array} \right)
\label{U2LS}
\end{equation}
where $\alpha = \tilde{a}$ or $\tilde{b}$ for (T = A, B).
In the time-evolution operators of Eqs.(\ref{UV}) and (\ref{U2LS}) there is an extra phase induced by the Stark-shifts, which has the same value as the dynamical phase induced by the pulse area.

Assuming that the absolute phases of all pulses are zero ($\phi_k = 0$),
excitation from $|11\rangle$ with a $\pi$-pulse prepares 
the entangled state $|\Psi_1\rangle = -\left( \tilde{a}_1 |r1\rangle + \tilde{b}_1|1r\rangle \right)$,
which can be de-excited by the second $\pi$ pulse. 
The essence of the SOP scheme is to find the spatial pulse parameters of the second pulse $\tilde{a}_2$, $\tilde{b}_2$ such that the prepared state after the first pulse $\ket{\Psi_1}$ is a dark state of $H_2^V$. 
We will now analyze in more detail the effect of the parameters by applying the time evolution operators of Eqs.(19) and (20).
When the system is initiated in the $|11\rangle$ state, the relevant
Hamiltonian for the sequence of the $3$ pulses is $H^V_1(t)h_1(t)+H^V_2(t)h_2(t)+H^V_3(t)h_3(t)$, where $h_k(t)$ are step functions defined in the time domain where the respective pulse acts. 
The total propagator is the ordered product of the corresponding propagators for each pulse (we assume the pulses act sequentially),
$U^{V} = \left( U^V_3 U^V_2 U^V_1 \right)$ (we omit subindexes for the total propagator).
Because we want to prepare a CZ gate (or in general a C-PHASE gate), 
the relevant matrix element of the propagator $U^V$ is the first element of the matrix, which gives the projection of the wave function on the $\ket{11}$ state after the pulse sequence. This must be $-1$, that is, the state must gain a $\pi$ phase after the time evolution.
After rearranging,
\begin{equation}
\begin{array}{ll}
U^V_{11} \equiv \expectation{11}{U^V}{11} & = 
e^{i (S_1+S_2+S_3)/2 }
\bigg\{ c_3 c_2 c_1 - e^{i(\phi_2-\phi_1)} (\vec{e}_2\vec{e}_1) c_3 s_2 s_1 
-e^{i(\phi_3 - \phi_2)} (\vec{e}_3\vec{e}_2) s_3 s_2 c_1    \\  & %
-e^{i(\phi_3 - \phi_1)} (\vec{e}_3\vec{e}_2)(\vec{e}_2\vec{e}_1) s_1 c_2 s_3 
-e^{i(\phi_3 - \phi_1)} \left[ \vec{e}_3\vec{e}_1 - (\vec{e}_3\vec{e}_2)(\vec{e}_2\vec{e}_1) \right] s_1 s_3
\bigg\}
\end{array}  \label{U11T}
\end{equation}
where, for brevity, we use a simplified notation with 
$c_k = \cos (S_k/2)$, $s_k = \sin(S_k/2)$

Clearly, the condition that $|\Psi_1\rangle = |\Phi_2^0\rangle$ (the dark state of ${\sf H}^V_2$) 
is that $\vec{e}_2$ is orthogonal to $\vec{e}_1$ and $\vec{e}_3$. 
For two-qubits, this
condition forces $\vec{e}_1$ to be parallel to $\vec{e}_3$.
Then Eq.(\ref{U11T}) becomes
\begin{equation} 
U^V_{11} =  
e^{i (S_1+S_2+S_3)/2 }
\bigg\{
\cos\theta_3 \cos\theta_2 \cos\theta_1  \mp e^{i(\phi_3 - \phi_1)} \sin\theta_3 \sin\theta_1 \bigg\}
\label{U11Ts}
\end{equation}
where the minus sign applies whenever $\vec{e}_3 = \vec{e}_1$ (hence, $\tilde{a}_3 = \tilde{a}_1$, $\tilde{b}_3 = \tilde{b}_1$)
and the plus sign when $\vec{e}_3 = -\vec{e}_1$ (hence, $\tilde{a}_3 = -\tilde{a}_1$, $\tilde{b}_3 = -\tilde{b}_1$).

As shown schematically before, the JP works perfectly under these conditions.
Making $S_1 = S_3 = \pi$ 
one obtains
$U_{11}^{V} = \mp e^{i (\pi + S_2/2 + \phi_3 - \phi_1)} = -1$ for $S_2 = 2\pi$
and $\phi_3 -\phi_1 = 0$ when $\vec{e}_3 = \vec{e}_1$, or $\phi_3 -\phi_1 = \pi$
when $\vec{e}_3 = -\vec{e}_1$.
In fact, one can regard the original protocol as a particular example of the orthogonality
of the geometrical factors, where $\vec{e}_1 = \vec{e}_3 = ( 1, 0 )$ and $\vec{e}_2 = ( 0, 1 )$.
Therefore, for the typical pulse areas, 
$S_1 + S_2 + S_3 = 4\pi$,
no phase adjustments in the pulses are needed to overcome the dynamical phase gained by the Stark shift. However, one can still play with $\phi_k$ in more general
set-ups.
The result is valid for any value of $\tilde{a}$ and $\tilde{b}$. 
And by adjusting the phase difference
$\phi_3 -\phi_1 = S_2/2 + \pi$, it 
also works for any value of the pulse area of the second pulse.

When the two-qubit system is initiated in $\ket{t}$ ($|01\rangle$ or $|10\rangle$), applying
Eq.(\ref{U2LS}), we obtain
\begin{equation}
U^T_{11} \equiv \expectation{t}{U^T}{t}  =
e^{i ( \alpha_1 S_1 + \alpha_2 S_2 + \alpha_3 S_3 ) /2}
\bigg\{
c_3 c_2 c_1 - e^{i(\phi_3-\phi_2)} s_3 s_2 c_1 - e^{i(\phi_3-\phi_1)} s_3 c_2 s_1
- e^{i(\phi_2-\phi_1)} c_3 s_2 s_1 \bigg\}
\end{equation}
where $c_k = \cos (\alpha_k S_k/2)$, $s_k = \sin (\alpha_k S_k/2)$,
$\alpha_k = \tilde{a}_k$ for $T = A$ and $\alpha_k = \tilde{b}_k$ for $T = B$.
The third pulse is a copy of the first pulse, so $S_3 = S_1$ and $\vec{e}_3 = \vec{e}_1$.
When $\phi_3 = \phi_2 = \phi_1$, 
\begin{equation}
U^T_{11} = 
e^{i ( \alpha_1S_1 + \alpha_2S_2/2)} \cos \left( \alpha_1 S_1 + \alpha_2 S_2/2 \right)
\end{equation}
Depending on the value of $\phi_k$ ($\phi_k = \phi_l \pm \pi$), all possible angle combinations can arise.
In the usual conditions, 
($S_1 = \pi$, $S_2 = 2\pi$)
\begin{equation}
U^T_{11} = e^{i ( \alpha_1 + \alpha_2) \pi} 
\cos \left( [\alpha_1 + \alpha_2] \pi \right)
\end{equation}
For orthogonal geometrical factors, 
$\tilde{a}_2 = \mp \tilde{b}_1 \equiv \mp \tilde{b}$, $\tilde{b}_2 = \pm \tilde{a}_1 \equiv \tilde{a}$,
$U_{11}^{A} = e^{i ( \tilde{a} - \tilde{b}) \pi}  \cos \left( [\tilde{a} - \tilde{b}] \pi \right)$,
$U_{11}^{B} = e^{i ( \tilde{a} + \tilde{b}) \pi}\cos \left( [\tilde{a} + \tilde{b}] \pi \right)$ or vice versa.

For independent qubits, $\tilde{a}=1$, $\tilde{b}=0$.  
Then $U^A_{11} = U^B_{11}$ 
and the fidelity
of the gate is $0$. (A different entangling gate, ${\cal C}^+$, is prepared with fidelity one, however.)
Otherwise, the real parts of $U_{11}^A$ and $U_{11}^B$, 
$\cos^2 \left([\tilde{a}-\tilde{b}]\pi\right)$ and
$\cos^2 \left([\tilde{a}+\tilde{b}]\pi\right)$ respectively, 
are always positive.
Therefore, regardless of the value of $\tilde{a}$ and $\tilde{b}$, it will not be possible to prepare the ${\cal C}^-$ gate with large fidelity without
controlling both the geometrical factors, $\vec{e}_k$, and the optical phases, $\phi_k$, as
will be shown numerically.

\section{Numerical results}

\begin{figure}[htb!]
\includegraphics[width=11cm]{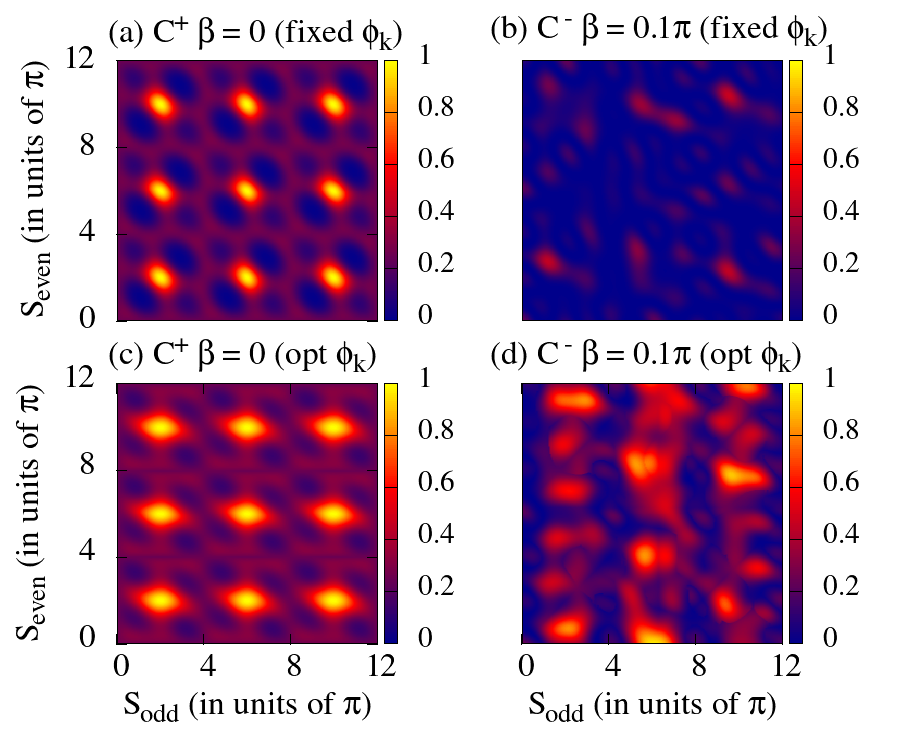}
\caption{
(a) Fidelity of the JP for the ${\cal C}^+$ gate with independent qubits for two-photon processes, as a function of the pulse areas in units of $\pi$. (b) Fidelity for the ${\cal C}^-$ gate when $\beta = 0.10\pi$. 
In (c) and (d) we show how the fidelity for the previous cases can be improved when the phases of the pulses $\phi_k$ are optimized for each value of the pulse areas.} 
\label{F10}
\end{figure}

As demonstrated analytically, the JP 
can prepare ${\cal C}^+$ gates with perfect fidelity when the effective two-photon pulses have areas $S_1 = S_3 = \pi$ while the second pulse has area $S_2 = 2\pi$. 
The minimal total pulse area is $4\pi$ in this case, but the solutions are periodic in the areas such that $S_1 + S_3 = (4m+2)\pi$ ($m = 0,1,\ldots \in \mathbb{Z}$) and $S_2 = (4m+2)\pi$ ($m \in \mathbb{Z}$).
This is shown in Fig.\ref{F10}(a), where we plot the fidelity of the gate as a function of the pulse areas. 
The results show that the set of protocols that implement the gate with high fidelity form a lattice~\cite{Sola_Nanoscale2023}.
To reveal more clearly the symmetries of the set of optimal protocols, we represent the fidelity as a function of the sum of the pulse areas of odd-numbered pulses, $S_\mathrm{odd} = S_1 + S_3$ and the sum of even numbered pulses, $S_\mathrm{even} = S_2$, which is just one pulse in this case. This choice 
of representation of the {\em fidelity map} will be more important when we consider pulse sequences with different number of pulses.

For a given pulse sequence defined by the pulse areas $S_1 = S_3$ and $S_2$ and the overlap angle $\beta = \arctan(\tilde{b}/\tilde{a})$, one can find the phases $\phi_k$ ($k=1,2,3$) that maximize the fidelity, for instance, using a Nelder and Mead simplex optimization scheme~\cite{Nelder_CJ1965,Powell_AcNum1998,Sola_PRA2023,Sola_AIPadv2023}. Applying this optimization to the $\beta = 0$ case, we show the resulting fidelity map in Fig.\ref{F10}(c). 
Although phase optimization is not strictly necessary to obtain
high-fidelity gates in the JP, 
it extends the solutions over a larger range of pulse areas.

On the other hand, we showed that the JP cannot prepare ${\cal C}^-$ gates with two-photon transitions. In fact, the maximum fidelity that can be achieved is $0.25$, which can be improved to $0.4$ by optimizing the pulse phases. 
Even if we allow non-independently addressed qubits with fixed phases (we showed results for $\phi_k = 0$, but similarly low fidelities were obtained for all other choices), the SOP scheme cannot implement the ${\cal C}^-$ gate, although the maximum fidelity improves slightly. 
For instance, one can gain a fidelity of $0.5$ with almost independent qubits, as given by a very small, but non-zero, overlap of $\beta = 0.045\pi$ ($\tilde{b}^2 = 0.02$).
If the phases are optimized, a maximum fidelity of $0.994$ can be achieved. 
Figs.~\ref{F10}(b) and (d) show the results for $\beta = 0.10\pi$ ($\tilde{b}^2 = 0.1$) 
with fixed and optimized phases, respectively. 

\begin{figure}[htb!]
\includegraphics[width=11cm]{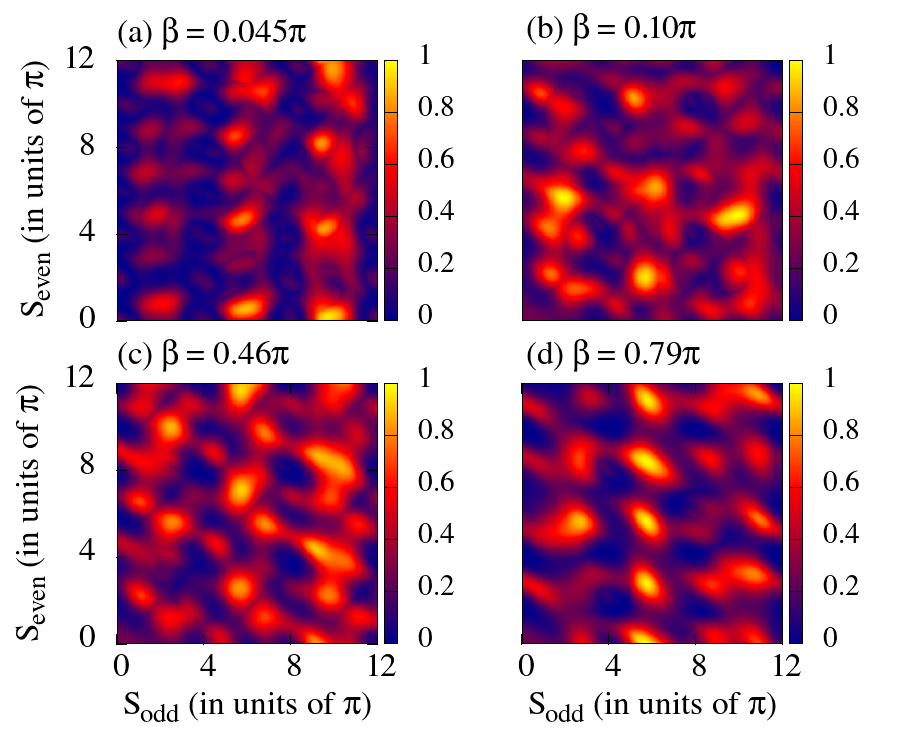}
\caption{Fidelity of the ${\cal C}^+$ gate as a function of the pulse areas for $3$-pulse sequences with (a) $\beta = 0.045\pi$ ($\tilde{b}^2 = 0.02$), (b) $\beta = 0.10\pi$ ($\tilde{b}^2 = 0.1$), (c) $\beta = 0.46\pi$ ($\tilde{b}^2 = 0.2$) and 
(d) $\beta = 0.79\pi$ ($\tilde{b}^2 = 0.5$). 
The pulse phases $\phi_k$ have been optimized for each value of the pulse areas.}
\label{Fvsb}
\end{figure}

The SOP performs better than the JP in preparing the ${\cal C}^+$ gate. While the highest fidelities are only slightly better (with values over $0.995$ for all tested angles), one can also observe larger regions in the parameter space defined by the pulse areas where one can find high-fidelity solutions, as shown in the fidelity map of Fig.~\ref{Fvsb} for different values of $\beta$.
The results are similar for any overlap angles, demonstrating the versatility of the SOP for different distances or arrangements of the qubits, which are encoded in the values of $\beta$. Once the phases are allowed to changed, both the positive and negative $\vec{e}_1 \pm \vec{e}_3$ conditions yield the same results after optimization. In fact, numerically one finds different sets of phases that maximize the fidelity for each value of the pulse areas or $\beta$.  
If the SOP conditions (orthogonality and symmetry) are relaxed and more parameters are allowed to be optimized, then the set of optimal protocols increases considerably~\cite{Sola_PRA2023,Sola_AIPadv2023}.
However, unlike in the single-photon SOP case, where the highest fidelities were mostly below $0.97$, optimizing the phases allows the fidelities of the two-photon version of the SOP to often exceed $0.99$.

\begin{figure}[htb!]
\includegraphics[width=11cm]{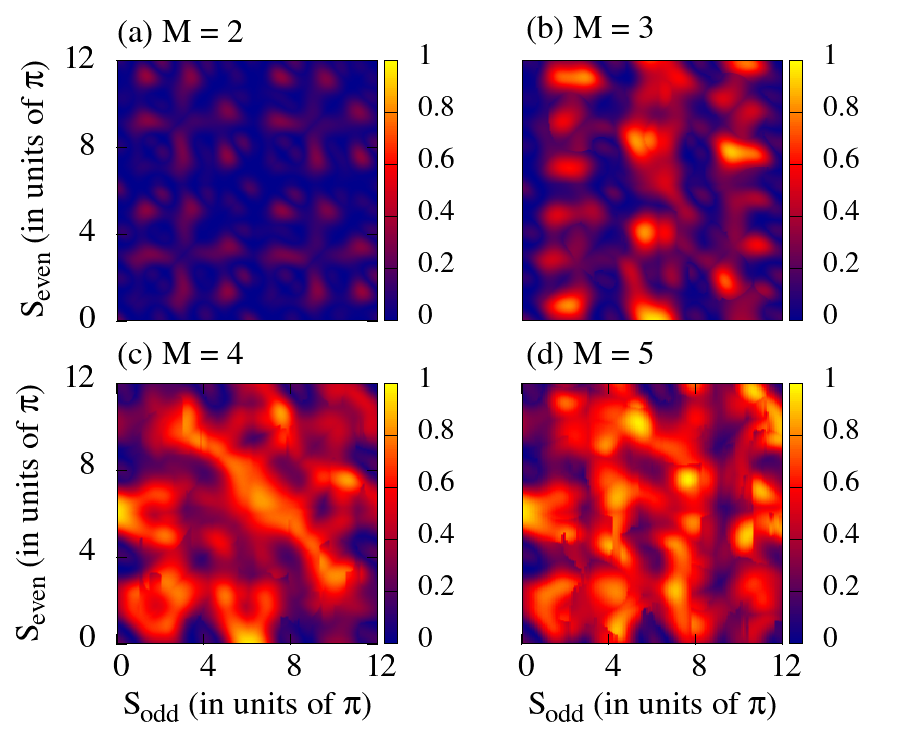}
\caption{Fidelity of the ${\cal C}^-$ gate as a function of the pulse areas for
pulse sequences with $\tilde{b}^2 = 0.1$ and 
(a) $M = 2$, (b) $3$, (c) $4$ and (d) $5$ pulses.
The pulse phases $\phi_k$ have been optimized for each value of the pulse areas.}
\label{FvsM}
\end{figure}

As in the single-photon case, the SOP scheme can be generalized to pulse sequences of arbitrary length. Here, we restrict our attention to highly symmetric sequences. In particular, all odd-numbered pulses ($k = 1, 3, \ldots$) share the same parameters (except for their phases), and all even-numbered pulses ($k = 2, 4, \ldots$) likewise share a common set of parameters. Moreover, odd and even pulses are mutually orthogonal, in the sense that $\vec{e}_k \cdot \vec{e}_{k+1} = 0$.
 
Specifically, for the simpler $2$-pulse sequences, the dynamics gives
\begin{equation}
    U^V_{11} = 
    e^{i ( S_1 + S_2 )/2}  \cos(S_1/2) \cos(S_2/2)
    \label{V2M}
\end{equation}
and
\begin{equation}
    U^T_{11} = 
    e^{i ( \alpha_1S_1 + \alpha_2S_2 )/2}  \cos\left( \alpha_1 S_1/2 +  \alpha_2 S_2 /2 \right)  \ .
    \label{TLS2M}
\end{equation}
These conditions restrict the search for parameters too severely. As follows from Eq.(\ref{V2M}), to maximize population return with a sign flip, the area of the pulses in the cosine terms should be even multiples of $\pi$ with a $2\pi$ difference between the areas.  
However, the prefactor
$\exp[i(S_1+S_2)/2]$
flips again the sign of the phase. Consequently, the Stark shift conspires against the phase shift, making it impossible to achieve $U^T_{11} \approx -1$ by controlling the phase of just two pulses. This will occur for both 
the ${\cal C}^+$ and ${\cal C}^-$ gates, or any entangling gate.
A maximum fidelity of $0.6$ can be achieved for different values of $\beta$. In contrast, fidelities closer to $0.9$ could be achieved in two-pulse sequences using the single photon version of the SOP~\cite{Sola_Nanoscale2023}.

\begin{figure}[htb!]
\includegraphics[width=11cm]{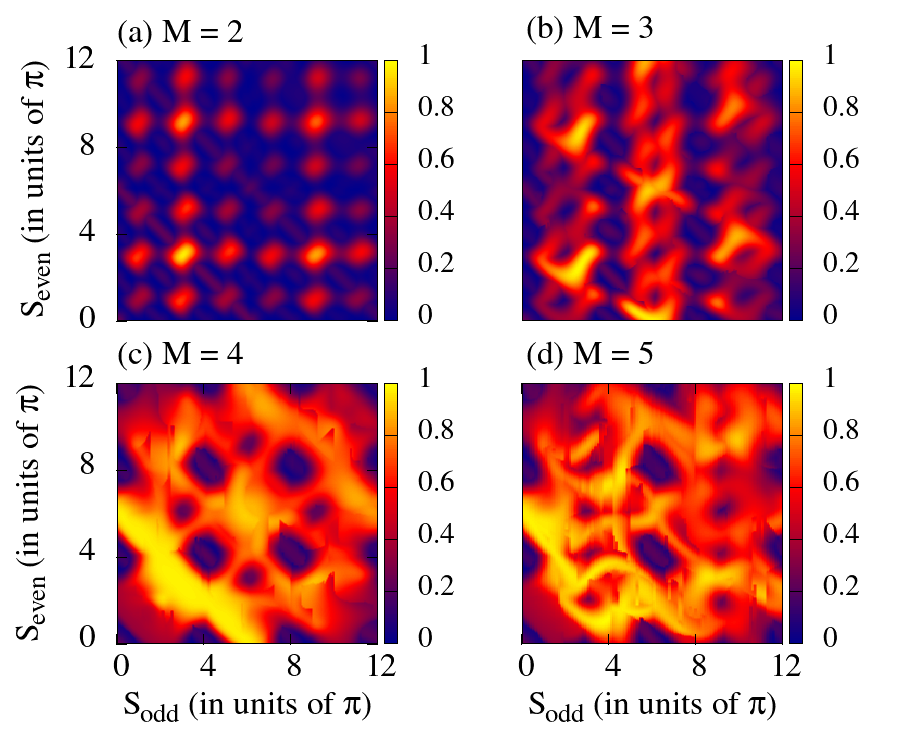}
\caption{Fidelity of the ${\cal C}^-$ gate as a function of the pulse areas implementing the SPP scheme for
pulse sequences with $\tilde{b}^2 = 0.1$ and 
(a) $M = 2$, (b) $3$, (c) $4$ and (d) $5$ pulses.
The pulse phases $\phi_k$ have been optimized for each value of the pulse areas.}
\label{FvsM-SPP}
\end{figure}

In Fig.~\ref{FvsM} we show the fidelity maps of the ${\cal C}^-$ gate using the SOP for different $M$-pulse sequences ($M = 2$ to $5$), fixing $\beta = 0.10\pi$ ($\tilde{b}^2 = 0.1$). Similar results, but with a higher density of high-fidelity protocols and slightly higher fidelities in the case $M\ge 3$, can be found for the ${\cal C}^+$ gate. 
For $M = 2$, the highest fidelities observed for pulse areas $S_\mathrm{odd}, S_\mathrm{even} \le 10\pi$ (and $\tilde{b}^2 = 0.1$) are $0.538$ for ${\cal C}^+$ and $0.322$ for ${\cal C}^-$. 
For $M = 3$, we found maxima of $0.995$
for ${\cal C}^+$ and $0.970$ for ${\cal C}^-$. 
For $M = 4$, these values increase to $0.999$ and $0.982$ for ${\cal C}^+$ and ${\cal C}^-$, respectively, and for $M = 5$, they reach $0.999$ and $0.994$ for ${\cal C}^+$ and ${\cal C}^-$.

A fair comparison of the performance of the different schemes requires to show the fidelities for the same integrated pulse area (the sum of the areas of all pulses), which is shown in the fidelity maps as a function of $S_\mathrm{odd}$ and $S_\mathrm{even}$.
In general, the density of optimal protocols increases with the number of pulses, correlated with the increase in the number of variational parameters (the absolute phases of the pulses, $\phi_k$) that are optimized. 
However, Fig.\ref{FvsM} reveals a similar number of bright spots
(very high fidelity protocols) for $4$-pulse sequences in comparison to $3$-pulse sequences, and significantly fewer than for $5$-pulse sequences.
The condition of orthogonality is unnecessarily restrictive, especially for pulse sequences with two pulses.
Alternatively, one can use the simpler {\em Symmetric Parallel Protocol} (SPP) where all structural vectors are identical, {\it i.e.} $\vec{e}_1 = \vec{e}_2 = \ldots$ (aligned or anti-aligned configurations yield 
the same results after phase optimization).
Then, for $2$-pulse sequences,
\begin{equation}
    U^V_{11} = 
    e^{i ( S_1 + S_2 )/2}  \bigg( \cos(S_2/2) \cos(S_1/2) \mp e^{i(\phi_2-\phi_1)} \sin(S_2/2) \sin(S_1/2) \bigg)
    \label{V2M-SPP}
\end{equation}
which, for $\phi_1 = \phi_2$, leads to $U^V_{11} =
\exp [i ( S_1+S_2 )/2] \cos\left[(S_2\pm S_1)/2\right]$,
while Eq.[\ref{TLS2M}] remains valid for the SPP case. The optimal parameters for these conditions must be approximate solutions of the Diophantine equations. These solutions
can be found for any value of $\beta$, but typically require larger pulse areas.

The fidelity maps for implementing the ${\cal C}^-$ gate with $\tilde{b}_1^2 = 0.1$ are shown in Fig.~\ref{FvsM-SPP} for the different pulse sequences.
We find maximum fidelities of $0.970$, $0.989$, $0.980$ and $0.996$ for $M = 2, 3, 4, 5$, respectively. Notably, high-fidelity gates can be achieved even with two-pulse sequences.

\begin{figure}[htb!]
\includegraphics[width=11cm]{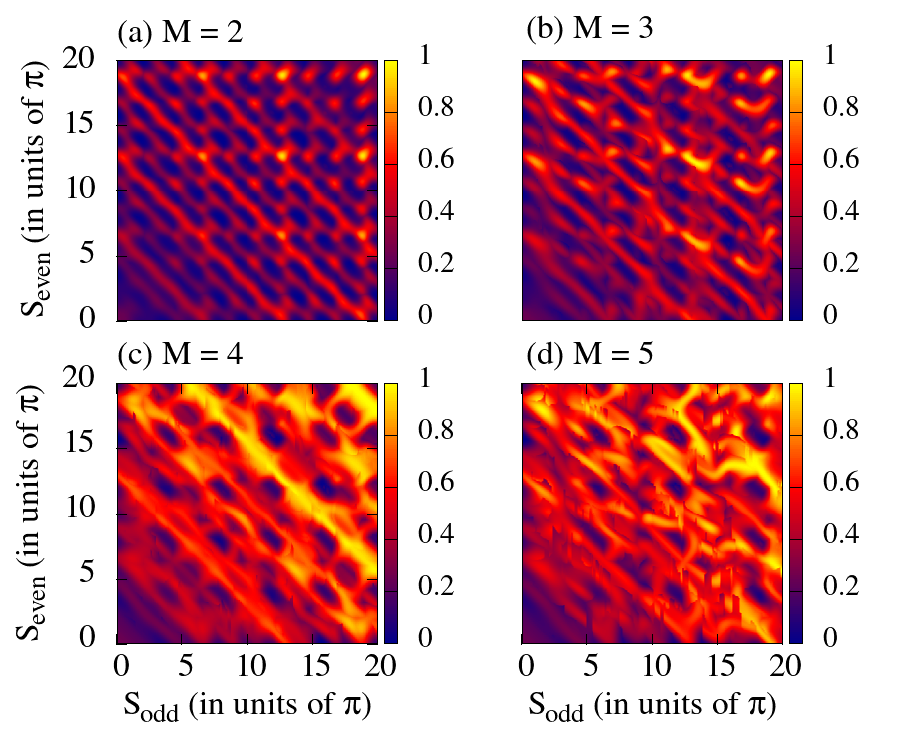}
\caption{Fidelity of the ${\cal C}^+$ gate as a function of the pulse areas implementing the SPP scheme for
pulse sequences with $\tilde{b}^2 = 0.1$ and 
(a) $M = 2$, (b) $3$, (c) $4$ and (d) $5$ pulses.
The pulse phases $\phi_k$ have been optimized for each value of the pulse areas.}
\label{FvsM-SPP2}
\end{figure}

Interestingly, significantly poorer results are obtained when the parameters are optimized to maximize the fidelity of the ${\cal C}^+$ gate in the SPP scheme, when the search is limited to
pulse areas ($S_\mathrm{odd}$ and $S_\mathrm{even}$) smaller than $10\pi$.
In Fig.~\ref{FvsM-SPP2} we show the fidelity maps with the range of areas extended up to $20\pi$.
Then we find nearly perfect gates (with gate errors smaller than $10^{-5}$) for even-pulse sequences, while for $3$-pulse sequences the highest fidelity is $0.901$. The optimal protocols typically align in strips, as analyzed for two-pulse sequences, in~\cite{Sola_PRA2024}.

Although the precise values for the highest fidelities depend on the choice of $\beta$, clearly the SPP performs better in the ${\cal C}^+$ gate only at larger pulse areas. Conversely, the SOP can perform better for smaller pulse areas, especially with odd-pulse sequences.
In any case, for sequences with more pulses, the number of parameters governing the dynamics is large enough to allow the implementation of high-fidelity gates for practically any pulse area above a certain threshold, using either the SOP or SPP scheme.

\section{Conclusions}

In this study, we identified a previously overlooked challenge in using the strong Rydberg blockade mechanism to create high-fidelity two-qubit phase gates. Dynamical phases from Stark shifts in non-resonant two-photon excitation can lower gate fidelity. We showed that the effective two-photon Hamiltonian changes the system’s evolution, making standard protocols like the Jaksch Protocol~\cite{Jaksch_PRL2000} and the Symmetrically Orthogonal Protocol~\cite{Sola_Nanoscale2023} unable to produce the desired $\mathcal{C}^-$ gate. Instead, without correction, these protocols naturally create a different entangling operation, the $\mathcal{C}^+$ gate.

We found that by controlling the optical phases of the pump and Stokes pulses, it is possible to solve this problem. Adjusting these phases helps reduce detrimental Stark-shift effects and enables high-fidelity entangling phase gates across a wide range of conditions. This led us to develop two new protocols for interacting qubits, called SOP and SPP. Their effectiveness depends on the length of the pulse sequence, which expands the possibilities for neutral-atom quantum computing. SOP works best with odd-length sequences and when optimizing the ${\cal C}^+$ gate. SPP is more effective for the ${\cal C}^-$ gate with smaller pulse areas. Using SPP to prepare the ${\cal C}^+$ gate can achieve perfect fidelity with larger pulse areas, especially for even-numbered pulse sequences.

While the results in this work were based on the validity of the adiabatic elimination, controlling the optical phases can be used to adjust the precise phases in numerically exact computations. Alternatively, two-photon Rydberg excitations can also be performed in resonant conditions limiting the population of the intermediate state~\cite{Carrasco_PRL2025}.

\section*{Acknowledgements}

This research was supported by Grant No. PID2021-122796NB-I00, 
funded by the MICIU/AEI/10.13039/501100011033 and the ERDF/EU, 
the MATRIX-CM project, TEC-2024/TEC-85, the DEVCOM Army Research Laboratory under Cooperative Agreement Number W911NF-24-2-0044
and by Grant No. NRF-2021R1A5A1030054, supported by the Center for Electron Transfer funded by the Korean government (MSIT). 
B.Y.C. acknowledges support from a María Zambrano grant
funded by the European Union - NextGenerationEU.

\bibliography{refs}

\end{document}